\documentstyle[numreferences]{kapproc}

\newcommand{\C}{\mbox{\boldmath$C$}}
\newcommand{\R}{\mbox{\boldmath$R$}}
\newcommand{\D}[1]{\,{\rm d}#1\,}
\newcommand{\im}{{\rm i}}

\begin{opening}
\title{Quantum Clifford Algebras}
\author{T. Brzezi\'nski%
\thanks{On leave from Institute of Mathematics, University if {\L}\'od\'z,
ul.~Banacha 22, 90--238 {\L}\'od\'z, Poland}}
\institute{Department of Applied Mathematics and Theoretical Physics\\
University of Cambridge\\
Silver St., CB3 9EW Cambridge, U.K.}
\author{L.C. Papaloucas}
\institute{Institute of Mathematics\\ University of Athens\\
 106 79 Athens, Greece}
\author{J. Rembieli\'nski}
\institute{Department of Mathematical Physics\\ University of {\L}\'od\'z\\
ul.~Pomorska 149/153, 90--236 {\L}\'od\'z, Poland}
\end{opening}

\begin{document}

\begin{abstract}
Quantum multiparameter deformation of real Clifford algebras
is proposed. The corresponding irreducible representations are
found.
\keywords Clifford algebras, quantum deformations
\end{abstract}

\section{Introduction}
As is well known there exists a direct relation between the
exterior and the Grassmann algebras. Let us
consider a unital algebra $C(x)$ over $\C$ freely
generated by the elements $x1,\ldots,xN$ and a two-sided ideal
$J_0\subset C(x)$ generated by $xixj-xjxi$,
with $i,j=1,\ldots,N$. Now, the
quotient algebra ${\bf M}_0=C(x)/J_0$ is freely generated by
$x1,\ldots,xN$ subject to the commutativity relation
\begin{equation}
xixj=xjxi.
\end{equation}
Introducing a new set of
generators $\!\D{xk}\!$, $k=1,\ldots,N$, where $\rm d$ is the exterior
differential operator, we can extend this algebra first to an
$A_0$-module and then to the exterior algebra $\Omega$, with the
standard product given by
\begin{eqnarray}
xi\D{xj}\!&=&\!\D{xj}xi,\\
\!\D{xi}\D{xj}\!&=&\!-\D{xj}\D{xi}\!.
\label{exterior}
\end{eqnarray}
The generators $\D xi$ define a finite dimensional subalgebra of
$\Omega$ with the multiplication given by the exterior product
(\ref{exterior}). This algebra is a differential realisation of the
abstract Grassmann algebra generated by the set $\hat\gammai$,
$i=1,\ldots , N$ subject to the relations
\begin{equation}
\hat\gammai\hat\gammaj+\hat\gammaj\hat\gammai=0.
\label{grassmann}
\end{equation}
Clifford algebra is defined as the central extension of the algebra
(\ref{grassmann})
\begin{equation}
\gammai\gammaj+\gammaj\gammai=2g{ij}I,
\end{equation}
with $g{ij}=g{ji}$, $\overline{g}{ij}=g{ij}$.

Now, it is easy enough to apply this procedure to the Manin's
hyperplane. In this case we choose the ideal $J$
\cite{manin} leading to the following reordering rules
\begin{equation}
xixk=q_{ik}xkxi,
\end{equation}
with $q_{ik}\in\C-\{0\}$, $q_{ik}=q{-1}_{ki}$, $q_{kk}=1$.
Consequently,  we obtain \cite{holomorph}
\begin{equation}
[B\!\D{xi}\D{xk}\!=-q_{ik}\D{xk}\D{xi}
\end{equation}
as basic rules for the two-form sector of the corresponding twisted
exterior algebra. Similarly as in the standard case,  identifying
$\!\D{xi}\!$ with $\hat{\mit\Gamma}i$, we obtain
\begin{equation}
\hat{\mit\Gamma}i\hat{\mit\Gamma}k+q_{ik}\hat{\mit\Gamma}k\hat{\mit\Gamma}i
=0,
\label{qgrassmann}
\end{equation}
as a generalisation of the Grassmann algebra multiplication rules.

In the next sections of the paper we construct central extensions of
the algebra (\ref{qgrassmann}).

\section{Clifford Algebra $C{p,q}$ in the Witt Basis}
The {\em canonical basis} $\{\gamma\mu\}$ for the
Clifford algebra $C{p,q}$ is defined as follows
\begin{equation}
\gamma\mu\gamma\nu+\gamma\nu\gamma\mu=2g{\mu\nu}I,
\label{cliff}
\end{equation}
where
\begin{equation}
g=\pmatrix{I_p&0\cr0&-I_q},
\end{equation}
$p\geq q$, $\mu,\nu=1,\ldots,(p+q)$.

For $p+q=2n$ there exists another standard basis, the so called {\em
Witt basis\/}
\cite{chevalley,spinor}. As we will see later this basis is suitable
for a non-commutative generalisation of the Clifford algebra. In the
Witt basis metric tensor takes the form
\begin{equation}
G=\pmatrix{0&I_n\cr I_n&0},
\end{equation}
while the new generators are defined by
\begin{eqnarray}
\gammaa_N=\frac{1}{\sqrt2}(\gammaa+\im\gamma{a+n}),&\quad&
\gammab_N=\frac{1}{\sqrt2}(\gammab-\gamma{b+n}),\\
\gammaa_P=\frac{1}{\sqrt2}(\gammaa-\im\gamma{a+n}),&\quad&
\gammab_P=\frac{1}{\sqrt2}(\gammab+\gamma{b+n}),
\end{eqnarray}
where $a=1,\ldots,(n-q)$, $b=(n-q+1),\ldots,n$.
Using Eq. (\ref{cliff}) we obtain the following commutation relations
\begin{eqnarray}
\gamma\alpha_N\gamma\beta_N&=&-\gamma\beta_N\gamma\alpha_N,\\
\gamma\alpha_P\gamma\beta_P&=&-\gamma\beta_P\gamma\alpha_P,\\
\gamma\alpha_N\gamma\beta_P&=&-\gamma\beta_P\gamma\alpha_N+
\delta{\alpha\beta}I,
\end{eqnarray}
$\alpha,\beta=1,\ldots,n$. In particular, we notice that
${\gamma\alpha_N}2={\gamma\alpha_P}2=0$.

We see that $\gamma\alpha_N$ and $\gamma\alpha_P$ span two
totally isotropic $n$-dimensional subspaces in the generating
sector of the Clifford algebra $C{p,q}$, $p+q=2n$.

The hermitian conjugation can always be chosen as \cite{montpelier}
\begin{equation}
{\gamma\mu}\dagger=g{\mu\mu}\gamma\mu ,
\end{equation}
so that
\begin{equation}
{\gamma\alpha_N}\dagger=\gamma\alpha_P.
\end{equation}
Therefore $\gamma\alpha_N$ ($\gamma\alpha_P$) behave like
{\em fermionic annihilation (creation) operators}.

For $p+q=2n+1$, it is necessary to add an extra generator
$\gamma0$, ${\gamma0}\dagger=\gamma0$ satisfying
\begin{equation}
{\gamma0}2=I,\quad
\{\gamma0,\gamma\alpha_N\}=\{\gamma0,\gamma\alpha_P\}=0.
\end{equation}

For $p+q=2n$, real forms $C{p,q}$ can be reconstructed
via
\begin{equation}
A=\sum{n-q}_{\alpha=1}
(z_\alpha\gamma\alpha_N+\overline{z}_\alpha\gamma\alpha_P)+
\sumn_{\beta=n-q-1}(a_\beta\gamma\beta_N+b_\beta\gamma\beta_P),
\end{equation}
where $z_\alpha\in\C$, $a_\beta,b_\beta\in\R$, while
for $p+q=2n+1$ we have
\begin{equation}
A'=A+a_0\gamma0,\quad a_0\in\R.
\end{equation}

\section{Quantum Deformation of Clifford Algebras $C{p,q}$}
According to the standard procedure we try to deform the
Clifford algebras in the Witt basis via the following Ansatz
\begin{eqnarray}
{\mit\Gamma}\alpha_N{\mit\Gamma}\beta_N&=
&-qN_{\alpha\beta}{\mit\Gamma}\beta_N{\mit\Gamma}\alpha_N, \qquad
{\mit\Gamma}\alpha_P{\mit\Gamma}\beta_P=
-qP_{\alpha\beta}{\mit\Gamma}\beta_P{\mit\Gamma}\alpha_P,\\
{\mit\Gamma}\alpha_N{\mit\Gamma}\beta_P&=
&-q_{\alpha\beta}{\mit\Gamma}\beta_P{\mit\Gamma}\alpha_N+
2\delta{\alpha\beta}{\mit\Delta},\\
{\mit\Delta}{\mit\Gamma}\alpha_N&=
&\xi\alpha_N{\mit\Gamma}\alpha_N{\mit\Delta},\qquad
{\mit\Delta}{\mit\Gamma}\beta_P =
\xi\beta_P{\mit\Gamma}\beta_P{\mit\Delta},
\end{eqnarray}
where ${{\mit\Gamma}\alpha_N}\dagger={\mit\Gamma}\alpha_P$,
${\mit\Delta}\dagger={\mit\Delta}$.

The parameters $qN_{\alpha\beta}$, $qP_{\alpha\beta}$,
$q{\alpha\beta}$, $\xi\alpha_N$, $\xi\alpha_P$ should satisfy
a number of conditions following from the consistency with
associativity and the hermitian conjugation rules. Solving these
constraints, we obtain (for
$q{N(P)}_{\alpha\alpha}\neq-1$)\footnote{The case
$q{N(P)}_{\alpha\alpha}=-1$ leads to deformations of the symplectic
or Crummeyrolle  Clifford algebra.}
\begin{eqnarray}
\xi\alpha_N&=&q{-1}_{\alpha\alpha},\quad
\xi\alpha_P = q_{\alpha\alpha}, \quad
\overline{q}_{\alpha\beta} = q_{\beta\alpha},\\
qN_{\alpha\beta}&=&q_{\alpha\alpha}/q_{\alpha\beta},\quad
qP_{\alpha\beta} = q_{\beta\beta}/q_{\alpha\beta},\quad
q_{\alpha\beta}q_{\beta\alpha} = q_{\alpha\alpha}q_{\beta\beta}.
\end{eqnarray}
Now, demanding the extension be {\em central\/}, we are led to
\begin{equation}
q_{\alpha\alpha}=1,
\end{equation}
so that
\begin{equation}
|q_{\alpha\beta}| = 1,\quad
qN_{\alpha\beta} = q_{\beta\alpha},\quad
qP_{\alpha\beta} = q_{\beta\alpha}.
\end{equation}
and $\mit\Delta$ can be chosen as $+I$ (notice that
$\gamma_N\gamma_P$ is normal, so positive definite).

Resulting is the following deformation of the Clifford
algebra, obtained  as a central extension of the $q$-deformed Grassmann
algebra:
\begin{eqnarray}
{\mit\Gamma}\alpha_N{\mit\Gamma}\beta_N&=
&-q_{\beta\alpha}{\mit\Gamma}\beta_N{\mit\Gamma}\alpha_N,\\
{\mit\Gamma}\alpha_P{\mit\Gamma}\beta_P&=
&-q_{\beta\alpha}{\mit\Gamma}\beta_P{\mit\Gamma}\alpha_P,\\
{\mit\Gamma}\alpha_N{\mit\Gamma}\beta_P&=
&-q_{\alpha\beta}{\mit\Gamma}\beta_P{\mit\Gamma}\alpha_N+
2\delta{\alpha\beta}I,
\end{eqnarray}
with ${{\mit\Gamma}\alpha_N}\dagger={\mit\Gamma}\alpha_P$ and
$\alpha,\beta=1,\ldots,n$, where
\begin{equation}
|q_{\alpha\beta}|=1,\qquad
\overline{q}_{\alpha\beta}=q_{\beta\alpha}.
\end{equation}

The odd case ($p+q=2n+1$) can be treated by the extending  the above
algebra by
\begin{equation}
{{\mit\Gamma}0}\dagger={\mit\Gamma}0,\qquad{{\mit\Gamma}0}2=I
\end{equation}
together with the Ansatz
\begin{equation}
{\mit\Gamma}0{\mit\Gamma}\alpha_N =
 -qN_\alpha{\mit\Gamma}\alpha_N{\mit\Gamma}0,\qquad
{\mit\Gamma}0{\mit\Gamma}\alpha_P =
 -qP_\alpha{\mit\Gamma}\alpha_P{\mit\Gamma}0.
\end{equation}
If we additionally demand the existence of the classical limit, we will
obtain
\begin{equation}
q{N(P)}_\alpha=1,
\end{equation}
i.e.\
\begin{equation}
{\mit\Gamma}0{\mit\Gamma}\alpha_{N(P)}=
-{\mit\Gamma}\alpha_{N(P)}{\mit\Gamma}0.
\end{equation}

\section{Representations}
 Here we construct a Fock space of  representations
of the deformed
Clifford algebra. We consider even and odd cases separately and we
conclude this section with the example of the deformed Dirac matrices in four
dimensions.

\subsection{The Even Case ($p+q=2n$)}
We define the vacuum state
\begin{equation}
{\mit\Gamma}\alpha_N|0\rangle=0,\qquad\langle0|0\rangle=1.
\label{vacuum}
\end{equation}
The basis of the Fock space can be defined via
\begin{equation}
|\sigma_1,\ldots,\sigma_n\rangle=
\left(\frac{{\mit\Gamma}n_P}{\sqrt2}\right){\sigma_n}\ldots
\left(\frac{{\mit\Gamma}1_P}{\sqrt2}\right){\sigma_1}|0\rangle,
\end{equation}
where $\sigma_\alpha=0,1$.

Consequently
\begin{eqnarray}
{\mit\Gamma}\alpha_N|\ldots,\sigma_\alpha,\ldots\rangle&=&
\sqrt2\delta_{\sigma_\alpha1}\prodn_{\beta=1+\alpha}
(-q_{\alpha\beta}){\sigma_\beta}|\ldots,0,\ldots\rangle,\label{N}\\
{\mit\Gamma}\alpha_P|\ldots,\sigma_\alpha,\ldots\rangle&=&
\sqrt2\delta_{\sigma_\alpha0}\prodn_{\beta=1+\alpha}
(-q_{\beta\alpha}){\sigma_\beta}|\ldots,1,\ldots\rangle.\label{P}
\end{eqnarray}

\subsection{The Odd Case ($p+q=2n+1$)}
When $p+q=2n+1$, the representations of $C{p,q}$ can be easily
derived from the representations of even $C{p,q}$ by the following
procedure. First, we replace the even case vacuum $|0\rangle$ by
$|0\pm\rangle$, defined by Eqs. (\ref{vacuum}) and
\begin{equation}
\Gamma0|0\pm\rangle=\pm|0\pm\rangle.
\end{equation}
The Fock space is generated from the vacuum by the actions of the
raising operators $\Gamma\alpha_P$, precisely as in the even case.
The action of $\Gamma\alpha_N$,  $\Gamma\alpha_P$ on a standard
state $|\sigma_1,\ldots,\sigma_n\pm\rangle$ is given by Eqs (\ref{N}-
\ref{P}) and
\begin{equation}
{\mit\Gamma}0|\sigma_1,\ldots,\sigma_n\pm\rangle=
\pm(-){\Sigma\sigma_\alpha}|\sigma_1,\ldots,\sigma_n\pm\rangle.
\end{equation}

\subsection{Example}
Let us consider the simplest non-trivial example: $n=2$, $p+q=4$,
$q_{12}=\kappa$, $|\kappa |=1$. Explicitly, we have the following
algebra
\begin{eqnarray*}
{\mit\Gamma}1_N{\mit\Gamma}2_N&=&-\kappa{-1}{\mit\Gamma}2_N
{\mit\Gamma}1_N, \qquad
{\mit\Gamma}1_P{\mit\Gamma}2_P = -\kappa{-1}{\mit\Gamma}2_P
{\mit\Gamma}1_P,\\
{\mit\Gamma}1_N{\mit\Gamma}2_P&=&-\kappa{\mit\Gamma}2_P
{\mit\Gamma}1_N, \qquad
{\mit\Gamma}2_N{\mit\Gamma}1_P = -\kappa{\mit\Gamma}1_P
{\mit\Gamma}2_N,\\
&&\{{\mit\Gamma}1_N,{\mit\Gamma}1_P\}=\{{\mit\Gamma}2_N,
{\mit\Gamma}2_P\} =2I.
\end{eqnarray*}
and
$$
({\mit\Gamma}1_N)2=({\mit\Gamma}2_N)2 =
({\mit\Gamma}1_P)2=({\mit\Gamma}2_P)2=0.
$$
According to the Eqs. ({\ref{N}--\ref{P}) we obtain the following
deformation of the Dirac matrices
\begin{eqnarray*}
{\mit\Gamma}1_P& = &\sqrt2\pmatrix{0&0&0&0\cr1&0&0&0\cr0&0&0&0
\cr0&0&-\kappa{-1}&0},\qquad
{\mit\Gamma}1_N=\sqrt2\pmatrix{0&1&0&0\cr0&0&0&0
\cr0&0&0&-\kappa\cr0&0&0&0},\\
{\mit\Gamma}2_P& =
&\sqrt2\pmatrix{0&0&0&0\cr0&0&0&0\cr1&0&0&0\cr0&1&0&0},\qquad
{\mit\Gamma}2_N=\sqrt2\pmatrix{0&0&1&0\cr0&0&0&1\cr0&0&0&0\cr0&0&0&0}.
\end{eqnarray*}

\section{Related Topics}
Let us conclude with pointing out a number of interesting problems.
One may investigate relations of the above twisted Clifford algebras
to:
\begin{itemize}
\item
multiparameter $q$-deformations of Spin and pseudo-orthogonal
groups (corresponding to quantum groups);
\item
$q$-spinors;
\item
exotic statistics (anyons, etc.).
\end{itemize}

\acknowledgements
This work is supported by the KBN grant No.\ 2 0218 91 01.
We would thank to University of Athens, Department of Mathematics for the
hospitality.

\end{document}